\documentclass[aps,prl,twocolumn,groupedaddress]{revtex4}

\usepackage{graphicx}
\usepackage{dcolumn}
\usepackage{bm}
\usepackage{color}

\begin{document}


\title{Band Jahn-Teller Instability
and Formation of Valence Bond Solid in a Mixed-Valent Spinel Oxide LiRh$_2$O$_4$}

%

\author{Yoshihiko Okamoto$^{1,*}$, Seiji Niitaka$^{1}$, 
Masaya Uchida$^{1}$, Takeshi Waki$^{2}$, Masashi Takigawa$^{2}$, 
Yoshitaka Nakatsu$^{3}$, Akira Sekiyama$^{3}$, Shigemasa Suga$^{3}$,
Ryotaro Arita$^{1,\dag}$ and Hidenori Takagi$^{1,4}$}
\affiliation{
$^{1}$RIKEN (The Institute of Physical and Chemical Research), Hirosawa 2-1, Wako 351-0198, Japan\\
$^{2}$Institute for Solid State Physics, 
	University of Tokyo, Kashiwanoha 5-1-5, Kashiwa 277-8581, Japan\\
$^{3}$Division of Materials Physics, Graduate School of Engineering Science,
Osaka University, Toyonaka 560-8531, Japan\\
$^{4}$Department of Advanced Materials, University of Tokyo, Kashiwanoha 5-1-5, 
Kashiwa 277-8581, Japan\\
}

\date{\today}

\begin{abstract}
We have synthesized a new 
spinel oxide LiRh$_2$O$_4$ with a mixed-valent configuration of Rh$^{3+}$ and 
Rh$^{4+}$.
At room temperature it is a paramagnetic metal, but on cooling, 
a metal-insulator transition occurs and a valence bond solid state is formed below 170 K.
We argue that the formation of valence bond solid is promoted by a band 
Jahn-Teller transition at 230 K and the resultant confinement of $t_{\mathrm{2g}}$ holes
within the $xy$ band. 
The band Jahn-Teller instability is also responsible for 
the observed enhanced thermoelectric power in the orbital disordered phase above 230 K.
\end{abstract}

\maketitle

Among the wide variety of structural categories of complex 
transition metal oxides, 
spinel, with chemical formula AB$_2$O$_4$, is one of the most common structures and 
provides a unique playground for the physics of geometrical frustration.
The B-sublattice of the spinel structure consists of a three-dimensional network of tetrahedra, 
known as the pyrochlore lattice and can underpin
strong geometrical frustration effects. 
When antiferromagnetically coupled spins are placed on the pyrochlore lattice, 
long range magnetic ordering is suppressed substantially 
leading to, amongst other things, 
quantum spin liquid behavior~\cite{QSL}.
In many cases, however, 
a nontrivial self-organized state of spins marginally emerges 
by means of coupling with lattice distortion and/or 
orbital ordering~\cite{ZnCr2O4}.
In analogy with spins, when a spinel B-sublattice is occupied by ions with a formally 
half-integer valence, suppression of charge ordering 
and the eventual formation of a nontrivial state of charges results. 
Such charge frustration has attracted much interest since the discovery of Verwey 
ordering in the spinel Fe$_3$O$_4$ (Fe$_1$Fe$_2$O$_4$) with 1 : 1 ratio of Fe$^{2+}$ and Fe$^{3+}$
on spinel B-sublattice several decades ago~\cite{Fe3O4-1}, 
the spatial pattern of the ordering of which has been the subject of much debate~\cite{Fe3O4-2}. 

Novel mixed-valent spinels have emerged recently, 
notably 
CuIr$_2$S$_4$~\cite{CuIr2S4-1},~\cite{CuIr2S4-2},
AlV$_2$O$_4$~\cite{AlV2O4-1},~\cite{AlV2O4-2} and 
LiV$_2$O$_4$~\cite{LiV2O4-1},~\cite{LiV2O4-2}.
CuIr$_2$S$_4$ and AlV$_2$O$_4$, with nominally 1 : 1 ratio of
Ir$^{3+}$ and Ir$^{4+}$ and V$^{2+}$ and V$^{3+}$, respectively,
were found commonly to show nontrivial charge ordering on cooling.
In the charge ordered state, an array of spin-singlet ``molecules'',
Ir octamer and V heptamer, were discovered, 
which might be viewed as a kind of valence bond solid (VBS)~\cite{CuIr2S4-2},~\cite{AlV2O4-2}.
In LiV$_2$O$_4$ with 1 : 1 ratio of V$^{3+}$ and V$^{4+}$, 
no charge or spin ordering takes place down to below 1 K resulting in a novel heavy Fermion
ground state being stabilised~\cite{LiV2O4-1},~\cite{LiV2O4-2}.
This may represent a charge analogue of quantum spin liquid.
LiV$_2$O$_4$ was recently discovered to show a charge ordering under pressure, 
where a VBS state analogous to CuIr$_2$S$_4$ and AlV$_2$O$_4$ is very likely 
formed~\cite{LiV2O4-Niitaka}.
The question arises whether such VBS state is common to this class of mixed-valent 
spinel oxides and what is the mechanism behind it.

\begin{figure}
\includegraphics[width=7.8cm]{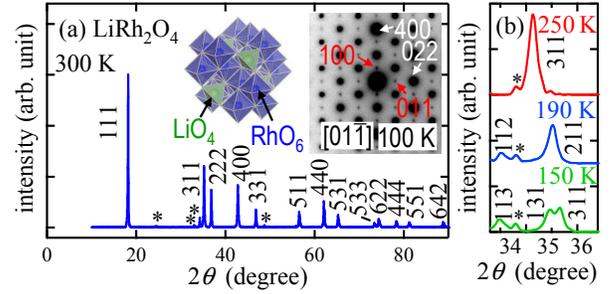}
\caption{\label{fig1}(color online) Powder XRD pattern of LiRh$_2$O$_4$ at 300 K (a) and the 
temperature evolution of cubic 311 reflection at 250 K, 190 K and 150 K (b).
The asterisks show reflections of Rh$_2$O$_3$ impurities.
(a) Reflections except for those asterisked could be indexed on the basis of the cubic
spinel structure with $a$ = 8.458 \AA.
The inset shows the crystal structure of LiRh$_2$O$_4$ in the cubic phase, 
spinel structure (left), and a [01\={1}] electron diffraction pattern at 100 K (right).
}

\end{figure}

\begin{figure}
\includegraphics[width=6.2cm]{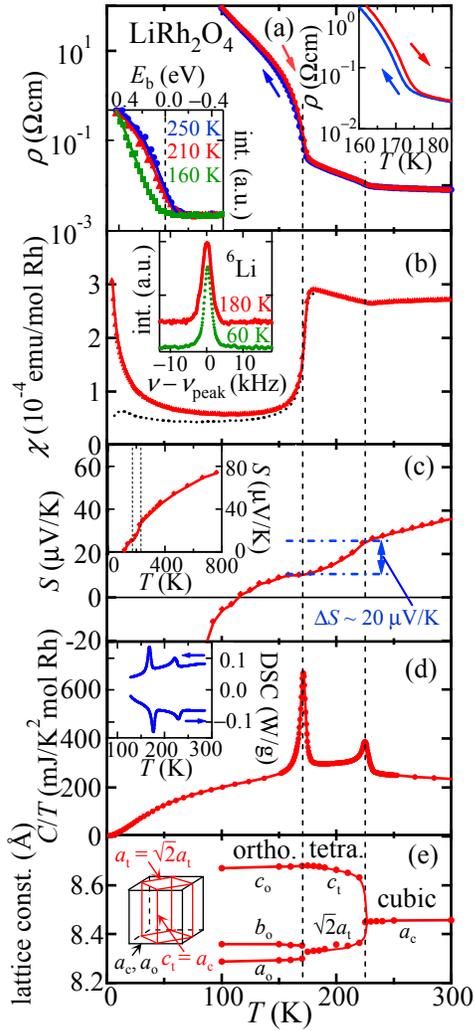}
\caption{\label{fig2}(color online) Temperature dependence of resistivity (a), magnetic susceptibility (b),
thermoelectric power (c), heat capacity divided by temperature (d)
and the lattice constants (e) of LiRh$_2$O$_4$ polycrystalline sample. 
(b) The dotted line shows the susceptibility after subtracting the Curie term 
with a Curie constant of 1.4 $\times$ 10$^{-3}$ emu/KmolRh.
Inset: (a) HAXPES spectra near the Fermi level at 
250 K, 210 K and 160 K (left) , 
and the resistivity around 170 K transition (right).
(b) $^6$Li-NMR spectra at 180 K and 60 K.
(c) Themoelectric power up to 800 K. 
(d) Differential scanning calorimetry curves obtained in heating (lower) and cooling (upper) process.
(e) A cubic, orthorhombic (black) and tetragonal (red) unit cell
of LiRh$_2$O$_4$. 
}
\end{figure}

In order to address these questions, we report
the discovery and the study of a mixed-valent spinel oxide LiRh$_2$O$_4$. 
LiRh$_2$O$_4$ consists of 1 : 1 ratio of low spin 
Rh$^{3+}$ ($S$ = 0, 4$d^6$) and 
Rh$^{4+}$ ($S$ = 1/2, 4$d^5$).
Rh$^{4+}$ has only one hole for six-fold orbital 
and spin degenerate $t_{\mathrm{2g}}$ orbitals and the coupling with these multiple
degrees of freedom may give rise to a channel to lift the degeneracy 
of charge distribution. 
We found that the ground state of this compound is 
a charge-ordered VBS triggered by a band Jahn-Teller transition.
This strongly suggests that orbital physics drives 
the VBS formation. 
In the non-Jahn-Teller phase at high temperatures, 
the proximity to the Jahn-Teller instability manifests itself as a 
drastic enhancement of thermoelectric power, providing us 
with a useful, new strategy to develop novel thermoelectric materials.

Ceramic samples of LiRh$_2$O$_4$ were synthesized by solid-state reaction.
Stoichiometric amounts of Li$_2$O$_2$ and Rh$_2$O$_3$ were mixed, 
and the mixture was calcined at 900$^\circ$C for 24 h under 5 atm. of oxygen pressure.
Sample characterization and structural analysis were performed by powder X-ray diffraction (XRD)
using Cu-K$\alpha$ radiation and electron diffraction.
Hard X-ray photoemission spectroscopy (HAXPES) measurements 
with an incident photon energy of 8175 eV were 
performed at the beamline BL19LXU in SPring-8.
$^6$Li-NMR spectra were obtained by Fast Fourier Transformation of the spin-echo signals.

The powder XRD pattern of LiRh$_2$O$_4$ at room temperature in Fig. 1 (a)
was consistent with the formation of a cubic spinel structure, 
where all the Rh sites are equivalent.
Figure 2 demonstrates the temperature dependence of various physical properties
of LiRh$_2$O$_4$.
The resistivity at 300 K is slightly less than 10 m$\Omega$cm and almost temperature
independent down to 230 K(Fig. 2 (a)). 
Since the sample is sintered polycrystal, empirically, 
the intrinsic resistivity 
can be more than one order of magnitude smaller than 10 m$\Omega$cm, 
consistent with the metallic nature.
In agreement with this, a well-defined Fermi cut-off was observed in the HAXPES
spectrum at 250 K, as shown in the inset of Fig. 2 (a).
The weakly temperature dependent paramagnetic susceptibility around 300 K can then be
interpreted as Pauli paramagnetism (Fig. 2 (b)).

With decreasing temperature, we observed a kink-like anomaly both 
in the resistivity and in the magnetic susceptibility at 230 K, suggestive of the second 
order phase transition. 
The resistivity shows a very weak increase 
on cooling down to 170 K.
As shown in the inset of Fig. 2 (a), the HAXPES spectrum at 210 K 
indicates the presence of a well-defined Fermi cut-off analogous to
those at 250 K.
It implies that in the temperature range between 170 K and 230 K 
the system remains poorly metallic and incoherent transport 
due to, for example, polaronic and/or disorder effects 
are likely responsible for the weakly semiconducting behavior of resistivity. 

At 170 K, the resistivity shows a discontinuous jump of several
orders of magnitude, 
suggestive of a metal-insulator transition, 
possibly associated with an ordering of charges (Fig. 2 (a)).
Simultaneously, a clear shift in the Fermi edge by $\sim$ 0.2 eV
to a higher binding energy side is observed in the 
HAXPES spectrum (inset of Fig. 2 (a)), 
indicating the opening of a charge gap with an energy scale of around a fraction of eV.
The presence of a tiny but clear hysteresis 
between warming and cooling
indicates that the transition at 170 K is first order in contrast to the transition 
at 230 K (inset of Fig 2 (a)).
Associated with the transition, the
magnetic susceptibility shows discontinuous drop
to a very small and almost temperature independent susceptibility,
implying the formation of a spin singlet.
The nonmagnetic nature of the insulating phase is firmly supported 
by $^6$Li-NMR measurement shown in the inset of Fig. 2 (b): 
the $^6$Li line width does not show any significant broadening 
in the insulating phase.
A thermally activated decay of the spin lattice relaxation rate 1/$T_1$,
with a large activation energy of $\sim$ 3000 K, was observed below 170 K, 
demonstrating the presence of a large spin gap~\cite{Waki}.
This implies that very robust singlet bonds are formed in the insulating
state, giving rise to the small magnetic susceptibility.
Note that the energy scale of the spin gap is comparable to that of 
the charge gap seen in the HAXPES spectrum.

\begin{figure}
\includegraphics[width=7.8cm]{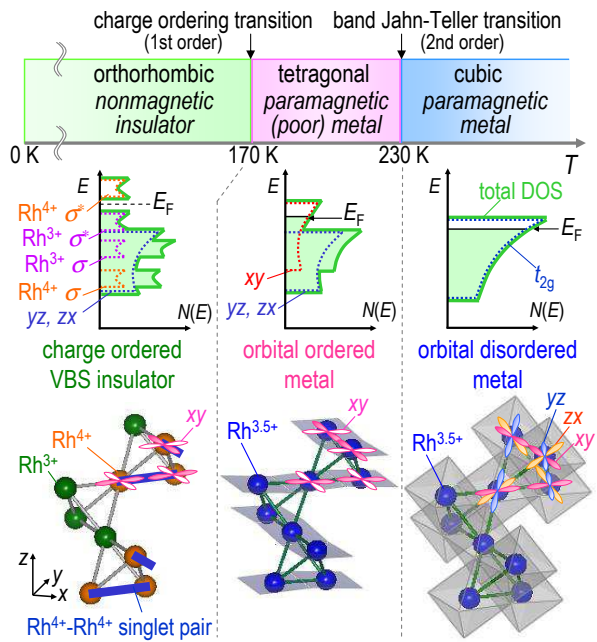}
\caption{\label{fig5}(color online)
Temperature variation of structural and electronic properties 
of LiRh$_2$O$_4$. 
A proposed model for electronic structure (DOS) evolution are also shown schematically (middle), together with the real space images of corresponding orbital states (lower). 
In the pictures of electronic structure, thick green lines and 
green-colored regions show the total DOS and the occupied
states of them, respectively.
Dotted lines show partial DOS of individual bands.
}

\end{figure}

The two transitions at 230 K and 170 K are accompanied by a
structural phase transition (Fig. 1 (b)). 
The diffraction pattern between 230 K and 170 K can be indexed with a tetragonal unit cell
($a_{\mathrm{t}}$ = $a_{\mathrm{c}}$/$\sqrt{2}$ = 5.889 \AA, $c$ = 8.665 \AA \ at 200 K).
Since we did not observe any extra spots in the electron diffraction compared with
those in the cubic phase, the cubic-tetragonal transition can be described as a simple
elongation of cubic spinel structure along [001] with space group $I$4$_1$/$amd$ (No. 141).
Note that the tetragonal distortion is surprisingly large with $c$/$a_{\mathrm{c}}$ = 
$c$/$\sqrt{2} a_{\mathrm{t}}$ = 1.04 at 180 K. 
The pattern in the insulating state below $T_{\mathrm{MI}}$ = 170 K can be indexed with
 an orthorhombic
cell ($a$ = 8.287 \AA, $b$ = 8.359 \AA, $c$ = 8.670 \AA \ at 100 K)~\cite{Ortho}.
In the electron diffraction patterns of the orthorhombic phase
($T < T_{\mathrm{MI}}$), 
new spots including 011 and 100 clearly emerge (inset of Fig. 1), 
indicative of a complicated lattice distortion and the resultant lowering of
symmetry with the charge ordering below $T_{\mathrm{MI}}$ = 170 K.
Electron diffraction and the synchrotron XRD pattern~\cite{PF} indicate that 
the new spots emerge at (2$l$ $+$ 1, 2$m$, 2$n$) and (2$l$, 2$m$ $+$ 1, 2$n$ $+$ 1),
implying that the propagation vector of modulation $\mathbf{k}$ = (1, 0, 0).
$\mathbf{k}$ of LiRh$_2$O$_4$
is distinct from those of CuIr$_2$S$_4$ ($\mathbf{k}$ = (0, 0, 1),
 (1/2, 1/2, 1/2))~\cite{CuIr2S4-2} 
and AlV$_2$O$_4$ ($\mathbf{k}$ = (1/2, 1/2, 1/2))~\cite{AlV2O4-1}.

The results of specific heat measurement shown in Fig. 2 (d) 
suggest that the transition at 230 K is in fact a drastic electronic transition
in addition to the 170 K transition.
The change of entropy $\Delta S_{\mathrm{entropy}}$ associated with the transitions, 
estimated from specific heat and DSC measurements,
was found to be 1.0 J/KmolRh at 230 K and 2.9 J/KmolRh at 170 K, 
which corresponds to 0.24$R$/mole Rh$^{4+}$ and 0.70$R$/mole Rh$^{4+}$, respectively. 
It may not be surprising to see a large entropy change of the order of $R$ 
at 170 K transition associated with a charge ordering.
It is surprising, however, to observe a large entropy change
for the cubic to tetragonal transition. 
A structural phase transition alone, as would have been expected from
the weak anomaly at 230 K in the resistivity and the magnetic susceptibility, 
is unlikely to be able to account for such a large entropy change. 
A drastic reconstruction of electronic states
should be invoked to account for the hidden and large entropy change.

The 230 K transition can be naturally understood now in terms of a 
band Jahn-Teller transition. 
The large tetragonal distortion ($c$/$a_{\mathrm{c}}$ $\sim$ 1.04) should split 
the triply degenerate $t_\mathrm{2g}$ band manifold into bands 
with stabilized $yz$ and $zx$ character and a band with destabilized $xy$ character (Fig. 3). 
Then, while the $yz$ and $zx$ bands are fully occupied with 4 electrons, 
the $xy$ band accommodates 0.5 holes (1.5 electrons). 
Stabilization of the fully occupied $yz$, $zx$ bands and 
destabilization of the partially filled $xy$ band enables the system to gain 
a Jahn-Teller energy. 
By inspecting the band structure of transition metal oxides with 
a spinel structure including LiTi$_2$O$_4$, LiV$_2$O$_4$ and ZnRh$_2$O$_4$,
we note that the transition metal $t_\mathrm{2g}$ band commonly has 
relatively high density of states (DOS) near the top of the band manifold, 
which arises from bands with flat dispersion produced by geometrical frustration~[15-18].
Recent band calculation indicated that the same is true for LiRh$_2$O$_4$~\cite{Arita}.
Since the Fermi level $E_\mathrm{F}$ should be located near the top of $t_\mathrm{2g}$ manifold 
in LiRh$_2$O$_4$, 
a relatively high DOS at $E_\mathrm{F}$ can be anticipated. 
Such high DOS can couple with the degenerate orbital degrees of 
freedom, leading to a band Jahn-Teller instability.
The large entropy change observed at 230 K then should be ascribed to 
the lifting of the band degeneracy through orbital ordering.

The high entropic state with band Jahn-Teller instability realized 
in the cubic phase 
can merit enhancing the thermoelectric power $S$.
The presence of a flat band near $E_\mathrm{F}$ 
can drastically enhance $S$, 
as pointed out for the well known thermoelectric Na$_x$CoO$_2$~\cite{terasaki}~\cite{NaxCoO2}.
LiRh$_2$O$_4$ indeed indicates a large and positive $S$ above 230 K, 
as large as 80 $\mu$V/K at 800 K,
where the system remains orbitally degenerate (Fig. 2 (c)).
Below the band Jahn-Teller transition at 230 K where the orbital degeneracy is lifted, 
$S$ rapidly decreases by $\sim$ 20 $\mu$V/K.
Assuming that $S$ is simply entropy per 
charge carrier $S_{\mathrm{entropy}}$/$ne$, 
the large entropy change $\Delta S_{\mathrm{entropy}}$ $\sim$ 1.0 J/KmolRh 
at the 230 K transition, estimated from the specific heat anomaly, 
gives an enhancement of thermoelectric power 
$\Delta S$ = 21 $\mu$V/K in the cubic (orbital disordered) phase, 
as compared with the tetragonal (orbital ordered) phase.
This agrees well with the observed change of 
thermoelectric power $\Delta S$ $\sim$ 20 $\mu$V/K, 
implying that the large entropy change at the cubic to tetragonal transition 
indeed represents an ``electronic'' reconstruction of degenerate 
bands and manifests itself as an enhanced $S$
in the cubic orbital-disordered phase.

Below 230 K, the conduction should be dominated by the carriers (0.5 holes)
in the band primary with $xy$ character. 
Since the $xy$ orbitals overlap strongly along the edge of Rh-tetrahedron 
perpendicular to the $c$-axis,
[110] and [1\={1}0], 
the $xy$ band
should be more dispersive along these directions and quasi-one-dimensional 
in nature (Fig. 3).
The confinement of Rh$^{4+}$ holes within the quasi-one-dimensional $xy$-band,
associated with the band Jahn-Teller transition, 
very likely promotes the formation of VBS despite 
the presence of charge frustration.
This picture is essentially nothing but the ``orbital induced Peierls'' transition 
proposed by Khomskii and Mizokawa as a model for the VBS formation 
in CuIr$_2$S$_4$~\cite{CuIr2S4-3}.
CuIr$_2$S$_4$ is isoelectronic to LiRh$_2$O$_4$ but shows only one 
transition from a paramagnetic metal to a spin-singlet insulator~\cite{CuIr2S4-1}. 
The lattice distortion in the insulating phase was decomposed by Khomskii and Mizokawa 
into Jahn-Teller like elongation of $c$-axis ($c$/$a$ $\sim$ 1.03) and 
tetramer formation within the [110] or [1\={1}0] chain 
and they argued that the transition is essentially a Peierls 
transition within the one-dimensional $xy$ band~\cite{CuIr2S4-3}.
The occurrence of a band Jahn-Teller transition  in LiRh$_2$O$_4$ at higher temperature than
the charge ordering transition clearly indicates that there 
is an intrinsic band instability in the 0.5 hole 
doped $t_\mathrm{2g}$ system on a frustrated pyrochlore lattice, 
and sets up the possibility of formation of VBS.
This implies that the band Jahn-Teller instability, 
rather than the instability for spin-singlet bond formation, 
is the driving force for the suppression of charge frustration in this system. 
In accord with this, although the Jahn-Teller distortion presents as a common 
ingredient of lattice distortion in the insulating state of LiRh$_2$O$_4$ and 
CuIr$_2$S$_4$, the overall lattice distortion pattern representing 
the charge ordering is clearly different between the two, 
implying the ordering pattern is determined reflecting materials-dependent details.

In conclusion, a new mixed-valent spinel oxide LiRh$_2$O$_4$ was synthesized, 
which provides us with an intriguing playground for the physics of charge-orbital-spin 
composites on geometrically frustrated lattices. 
A novel interplay of the orbital degeneracy of $t_{\mathrm{2g}}$ manifold and 
the high electronic density of states, linked to the unique geometry 
of the pyrochlore
lattice, 
gives rise to a band Jahn-Teller instability in this 0.5 hole-doped system. 
This drives the system to a transition from an orbital-disordered metal 
to an orbital-ordered metal at 230 K. 
The large electronic entropy in the orbital disordered phase manifests itself 
as a remarkable enhancement of thermoelectric power, 
compared with the orbital-ordered metal. 
Materials displaying a band Jahn-Teller instability could be 
a good strategy to develop
high performance thermoelectric materials. 
A charge-ordered VBS state is formed below 170 K. 
The formation of VBS is apparently driven by 
the band Jahn-Teller transition, which confines the holes within the 
quasi-one-dimensional $xy$ band. 
Such a charge ordering induced by an orbital ordering appears to be 
quite ubiquitous to this class of charge frustrated mixed-valent systems. 

We thank Z. Hiroi, H. Kuriyama, J. Matsuno, M. Nohara, R. S. Perry, 
H. Sawa, N. Shannon, A. Yamamoto, K. Kuroki, K. Held, A. V. Lukoyanov,
S. Skornyakov and V. I. Anisimov
for stimulating discussion, and A. Higashiya, M. Yabashi, K. Tamasaku, 
T. Ishikawa, S. Imada, H. Fujiwara, M. Yano, J. Yamaguchi and
G. Funabashi for supporting the HAXPES measurement.
A. S. acknowledges the support from the Asahi Glass Foundation.
M. U. acknowledges the support 
by ``Nanotechnology Support Project'' of the
Ministry of Education, Culture, Sports, Science and Technology (MEXT), Japan.
This work was partly supported by a Grant-in-Aid for Scientific Research on Priority Areas
``Novel States of Matter Induced by Frustration'' (No. 19052008), Japan.

\newpage
\clearpage

\textbf{Supplementary information.}

\textbf{Comment on the crystal structure of the charge-ordered insulating phase.}
Below 170 K, LiRh$_2$O$_4$ shows complicated lattice distortion due to the 
dimerization of Rh$^{4+}$.
We are able to index the X-ray diffraction pattern at 100 K with an orthorhombic unit cell.
Since some peaks originating from the orthorhombic phase, however, show a noticable 
broadening, there might be a further lowering of symmetry.
The superlattice spots including 100 and 011 in fact appears in a 
[01\={1}] electron diffraction pattern.
We determined a reciprocal lattice of the orthorhombic phase as shown in the supplementary figure
(a), by combining it with the electron diffraction patterns along some other directions
and the synchrotron X-ray diffraction pattern.
This reciprocal lattice indicates that the propagation vector of modulation is 
$\mathbf{k}$ = (1, 0, 0).
On the other hand, Rh$^{4+}$-Rh$^{4+}$ dimers are likely formed in the $xy$ plane,
because the $xy$ band is preferentially occupied by 0.5 holes in the tetragonal phase, 
which is also supported by the lattice distortion and 
Ir$^{4+}$-Ir$^{4+}$ dimerization 
pattern in spinel sulfide CuIr$_2$S$_4$ isoelectronic to LiRh$_2$O$_4$.
Considering that the charge ordering modulated by $\mathbf{k}$ = (1, 0, 0) and 
the Rh$^{4+}$-Rh$^{4+}$ dimerization in the $xy$ plane, 
the only two charge-ordering configuration models,
shown in the supplementary figure (b), are allowed.

\begin{figure}
\includegraphics[width=7cm]{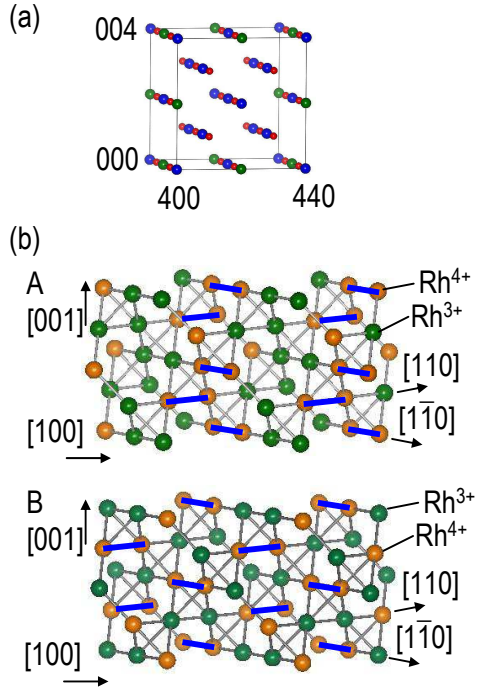}
\caption{\label{figS}
Supplementary figure. (a) A schematic reciprocal lattice of 
the orthorhombic phase of LiRh$_2$O$_4$ determined 
by electron diffraction measurements (crystal axis was determined by 
the synchrotron X-ray powder diffraction).
Blue and green spheres represent fundamental spots of the cubic spinel structure and
the forbidden spots, which were observed at both 300 K of the cubic phase 
and at 100 K of the orthorhombic phase due to multiple reflections, respectively.
Small red spheres represent new spots at (2$l$ + 1, 2$m$, 2$n$) and
(2$l$, 2$m$ $-$ 1, 2$n$ $-$ 1), where $l$, $m$ and $n$ are interger numbers, 
only observed at 100 K of the orthorhombic phase.
(b) Pictures of two charge ordering configuration models suggested by
the reciprocal lattice (a) under condition 
that all Rh$^{4+}$ ions form singlet bonds 
within the chains, represented by thick blue lines. 
Both models consist of Rh tetramers with an alternation of 
Rh$^{4+}$-Rh$^{4+}$-Rh$^{3+}$-Rh$^{3+}$ along [110] or [1\={1}0] Rh chain.
}
\end{figure}


\begin{thebibliography}{199}
\bibitem{Ortho}
		Since some peaks show a noticable 
		broadening, there might be a further lowering of symmetry.
\end{thebibliography}
\end{document}